# A GRASPxELS with Depth First Search Split Procedure for the HVRP

**Christophe Duhamel**[1]
**Lacomme Philippe**[1]   **Caroline Prodhon**[2]

[1] Université Blaise Pascal
Laboratoire d'Informatique (LIMOS) UMR CNRS 6158,
Campus des Cézeaux, 63177 Aubiere Cedex
placomme@sp.isima.fr, duhamel@isima.fr

[2] Université de Technologie de Troyes,
ICD (équipe LOSI) UMR CNRS 6279
12, Rue Marie Curie, BP 2060, F-10010 Troyes Cedex (France)
caroline.prodhon@utt.fr

# Abstract

Split procedures have been proved to be efficient within global framework optimization for routing problems by splitting giant tour into trips. This is done by generating optimal shortest path within an auxiliary graph built from the giant tour. An efficient application has been introduced for the first time by Lacomme *et al.* (2001) within a metaheuristic approach to solve the Capacitated Arc Routing Problem (CARP) and second for the Vehicle Routing Problem (VRP) by Prins (2004). In a further step, the Split procedure embedded in metaheuristics has been extended to address more complex routing problems thanks to a heuristic splitting of the giant tour using the generation of labels on the nodes of the auxiliary graph linked to resource management. Lately, Duhamel *et al.* (2010) defined a new Split family based on a depth first search approach during labels generation in graph. The efficiency of the new split method has been first evaluated in location routing problem with a GRASP metaheuristic. Duhamel *et al.* (2010) provided full numerical experiments on this topic.

This research report focuses on the application of Depth First Search Split strategy for Heterogeneous VRP (HVRP) trying to extend the Split proposals introduced on the LRP to a new routing problem. The numerical experiments use classical HVRP instances and a set of new real life instances matching to the French districts.

*Keywords: Split, HVRP*

# Résumé

Les procédures de type Split ont été prouvées particulièrement efficaces en application au cœur de méthodes globales d'optimisation pour les problèmes de tournées. Le principe général de ces procédures consiste à découper une tournée géante en tournées réelles. Ceci se fait généralement en calculant optimalement un plus court chemin dans un graphe auxiliaire réalisé à partir de la tournée géante. Une application efficace a été introduite pour la première fois au sein d'une métaheuristique lors d'une résolution dédiée au problème de tournées sur arcs (CARP) par Lacomme *et al.* (2001) et ensuite pour le problème de tournées de véhicules (VRP) par Prins (2004). Plus tard, une version étendue de la procédure Split a été développée afin de traiter des problèmes de tournées encore plus complexes en se basant sur un découpage heuristique du tour géant grâce à une génération de labels dans le graphe auxiliaire permettant ainsi la gestion de ressources. Enfin dernièrement, Duhamel *et al.* (Lacomme *et al.*, 2010) ont défini une nouvelle famille de Split basée sur une exploration des labels du graphe en profondeur d'abord. L'efficacité de cette nouvelle génération de Split a d'abord été évaluée sur le problème de localisation-routage (LRP) résolu par une métaheuristique de type GRASP. Duhamel *et al.* (Lacomme *et al.*, 2010) ont fourni des résultats expérimentaux complets sur ce sujet.

Ce rapport de recherche se focalise plus particulièrement sur une application du Split utilisant la stratégie de recherche en profondeur d'abord pour le problème de tournées de véhicules à flotte hétérogène (HVRP). Le but est d'étendre la nouvelle stratégie de Split introduite pour le LRP à une autre classe de problèmes de tournées. Les tests expérimentaux utilisent des instances classiques du HVRP ainsi qu'un nouveau jeu d'essais ayant une configuration plus réaliste s'appuyant sur la géographie des départements français.

***Mots-clés : SPLIT, HVRP***

# 1 Introduction

The most famous routing problem is the Traveling Salesman Problem (TPS) where a single vehicle can visit all the clients (ref). When the capacity of one vehicle cannot supply the whole demand, it leads to the Vehicle Routing Problem (VRP). The latter can be formally defined on a complete, weighted and directed network with $n+1$ nodes. The depot is represented by node 0 and the customers by nodes from 1 to $n$. An unlimited fleet of identical vehicles with a capacity $Q$ is available to serve the demand $d_j$ of each customer $j$ from the depot. $D = \sum_{j \in J} d_j$ denotes the total demand. The cost to travel from node $i$ to node $j$ is $c_{ij}$. A solution of the problem consists in designing a set of trips of minimal cost to serve the customers. The following constraints must be taken into account:

- deliveries cannot be split (each customer must be served by a single vehicle);
- each route starts and ends at the depot;
- the total demand of the customers served by one vehicle must fit its capacity.

The VRP is NP-hard and exact methods experience strongly large computational time to handle instances with up to 100 customers. Therefore, heuristic approaches are the only mean to tackle large instances in acceptable computation time. However, the VRP may not be realistic enough for some companies. Indeed, when $K > 1$ types of vehicles are available, the model needs to be generalized. That is what is proposed in the Vehicle Fleet Mix Problem (VFMP) introduced by Golden *et al.* (1984). Each type $k$ has specific capacity $Q_k$ and fixed cost $f_k$. Choi and Tcha (2007) add a cost per distance unit $v_k$. The goal remains the same as for the VRP except that the total cost of a trip of length $L$ is $f_k + L.v_k$. If in addition, the company has already bought its vehicles, each type $k$ has a limited number $a_k$ of vehicles which introduces new constraints in VFMP_V instances which does not favor heuristic and metaheuristic resolution scheme. More precisely, the problem becomes the Heterogeneous Fleet VRP (HVRP). These extensions are NP-hard since they reduce to the VRP when $K = 1$ and $a_k = n$. For a fine review of the literature on VFMP and HVRP, see the recent publication from Prins (2009) where he proposes two memetic algorithms which tackle for the first time each kind of non-homogeneous fleet VRP, that is to say VFMP with Fixed cost per vehicle type only (VFMP-F), VFMP with Variable cost per distance unit only (VFMP-V), VFMP with Fixed and Variable costs (VFMP-FV) and HVRP.

This paper deals with the HVRP solved heuristically by a hybridation between a Greedy Randomized Adaptive Search Procedure (GRASP) and an Evolutionary Local Search (ELS), whose principle is explained in Section 2. The key feature of the proposed method is based on the search space investigation divided into two parts: giant tour space and whole solution space. The relation from the former to the later is made by a splitting procedure. Classical greedy split principle is reminded in Section 3. This split version is match up to a new split procedure denoted depth first search split (DFS), detailed in Section 4. Numerical experiments on Section 5 show the performances of the GRASPxELS with both split procedures.

## 2    GRASP$\times$ ELS framework

### 2.1 Key features

The purpose of this section is to evoke the principles of GRASPxELS where:

- GRASP (Greedy Randomized Adaptive Search Procedure) is a multi-start local search metaheuristic in which each initial solution is constructed using a greedy randomized heuristic and then improved by local search (Feo and Resende, 1995).
- ELS (Evolutionary Local Search) is an evolved version of ILS (Iterated Local Search). Starting from an initial solution, each ILS iteration consists in taking a copy of the solution *S*, applying a mutation operator, and improving the mutated solution by a local search. The resulting solution *S'* becomes the incumbent solution S. The ELS, introduced by Prins (2009b) for routing problems, is similar but, at each iteration *nd* "children" instead of 1 are generated from S, using mutation and local search, and the best child replaces S.

GRASP$\times$ELS (Prins, 2009b) is a hybridization of both GRASP and ELS capturing the positive features of both methods (figure 1). The initial solution iteratively generated by a greedy randomized heuristic within the GRASP aims at managing diversity in search space investigation while intensification occurs during the local search step hybridized with the ELS to better investigate the current local optimum neighborhood, before leaving it.

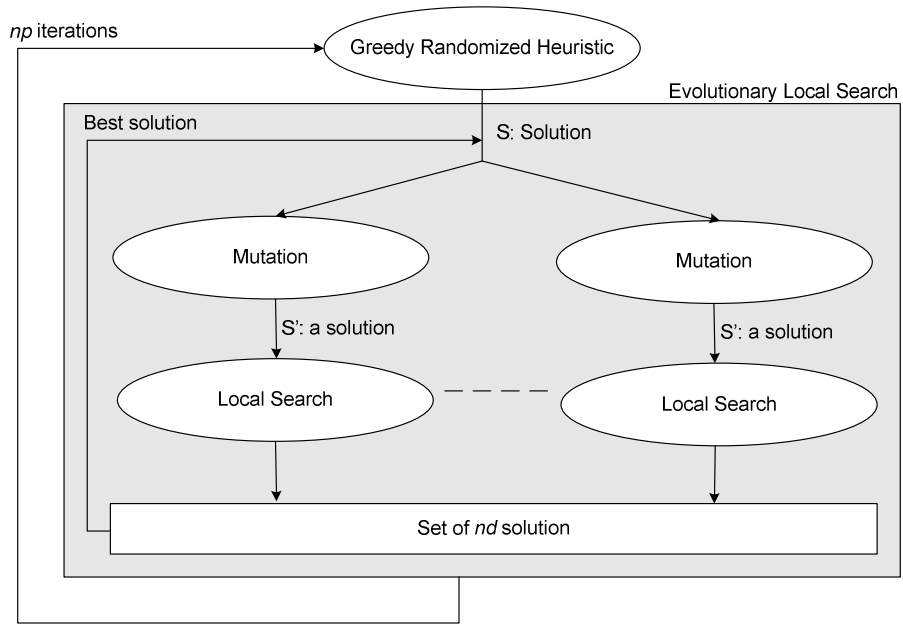

*Figure 1. GRASPxELS*

## 2.2 Search Space investigation strategy

An efficient search strategy for routing problems is based on a swap between two solution representations: solutions encoded as giant tours (TSP tours on the n customers) and solutions encoded as the set of trips (figure 2). Such an approach allows to work on the giant tour space (which is smaller than the space of solutions) before focusing on routing solutions. To obtain a solution, a giant tour is $T$ is converted by a procedure called Split into a solution S. A *Concat* procedure converts $S$ into a giant tour $T'$ by concatenating its trips. Split has been successfully applied to numerous routing problems including the Capacitated Arc Routing Problem (Lacomme *et al.*, 2001), the Vehicle Routing Problem (Prins, 2009a), the Location Routing Problem (Duhamel *et al.*, 2009). The high quality solutions obtained by Christian Prins (2009a) for the VRP and its extension, alternating between two search spaces (giant tours and routing solutions) push numerous researchers into promoting this line of research.

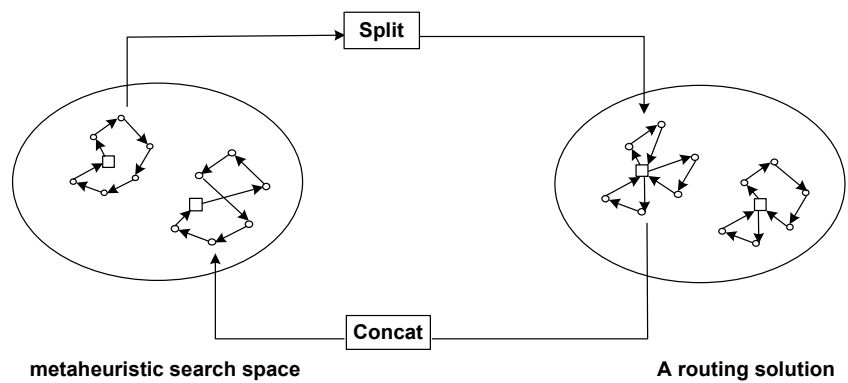

*Figure 2. Combination of the two search space*

## 2.3 GRASP×ELS framework for heterogeneous vehicle routing problems

To solve the HVRP, the GRASP×ELS process uses the specific search space investigation strategy based on Split as explain in Section 2.2.

The initial solutions of the GRASP scheme are generated using two procedures: `Heuristic_Generation_of_initial_solution` and `Random_Generation_of_initial_solution`. `Heuristic_Generation_of_initial_solution` is a greedy randomized constructive heuristic that first build a giant tour where a given node j is added after the previous one *i* on the nearest neighbor principle. To convert the giant tour $T$ into a HVRP solution S with respect to the HVRP specificities *i.e.* the availability of vehicles, the Split procedure has to tackle resource constraints. The obtained solution is improved by local search on the trips. The *Concat* procedure converts $S$ into a giant tour $T$ by concatenating its trips. The heuristic is stated as randomized since the node *j* to insert is randomly selected under a set of the *k* nearest nodes of *i*. The second heuristic, `Random_Generation_of_initial_solution`, is a fully random generation of giant trips which encompass a split of the giant tour and a local search on the solution. These two procedures can fail to find a HVRP solution depending on the giant tour. Indeed, the vehicle fleet size and vehicle capacity are constraints which can be responsible of numerous unsuccessful split attempts. The Local Search mentioned above is a first improvement descent method using several classical VRP neighborhoods to improve the initial solution. It is limited to `nls` iteration per call.

Algorithm 1 presents an overview of the method. $T$, $S$ and $f(S)$ respectively denote a giant tour, a HVRP solution and its cost. During the algorithm, the incumbent solution is stored in $S^*$ and $f^*$ is its value. Lines 15-54 correspond to the GRASP×ELS loop. It generates $np$ pairs $(S;T)$ used as starting points by the embedded ELS. Since as explain above, because of the resources constraints the Split procedure can fail to find a HVRP solution depending on the giant tour, a maximum number of `ni` calls to `Heuristic_Generation_of_initial_solution` is set (lines 17-20). If the greedy solution generation failed (`i>ni`), the random solution generation is used to obtain a solution at lines 23-26. For the same reason, only `niv` attempts are allowed in generation of neighbors, as shown on line 42 of algorithm 1.

The $ns$ iterations of ELS are performed in the loop lines 34-53 and the $nd$ mutations are completed in lines 36-48. A mutation is a random perturbation of the giant tour sequence.

```
1.   procedure GRASP_xELS
2.   global parameters
3.     np: number of GRASP iterations (initial solutions)
4.     ns: maximum number of iterations per ELS
5.     nr: maximum number of iterations without improvement per ELS
6.     nd: number of diversifications (mutations)
7.     nls: number of iteration assign to local search
8.     ni: maximum number of attempts in generation of initial heuristic solution
9.     niv: maximum number of attempts in generation of neighbords
10.    p : gap in percent to make intensification
11.  output parameters
12.    S*: best 2L-CVRP solution found
13.  begin
14.    f* := ∞; O := ∅
15.    for m := 1 to np do
16.      i:=0
17.      repeat
18.        (S,T) := call Heuristic_Generation_of_initial_solution ()
19.        i:=i+1
20.      until (S is a feasible solution) or (i> ni)
21.      if (i>ni) then
22.        i:=0
23.        repeat
24.          (S,T) := call Random_Generation_of_initial_solution ()
25.          i:=i+1
26.        until (S is a feasible solution)
27.      endif
28.      if (f(S) < f*) then
29.      f* := f(S); S* := S endif
30.      if (f(S)>p.f*) then
31.          S := S*;                    // intensification
32.      endif
33.      i, r := 0
34.      while (i < ns) and (r < nr) do   // ELS loop
35.        i := i + 1; f" := ∞
36.        for j := 1 to nd do            // mutation loop
37.          k:=0
38.          repeat
39.            T' := call Mutation (T)
40.            S' := call Split (T')
41.            k:=k+1
42.          until (S' is a solution) or (k> niv)
43.          if (S' is a solution) then
44.            S' := call Local_Search (S')
45.            T' := call Concat (S')
46.          endif
47.          if (f(S') < f") then f" := f(S'); T" := T'; S" := S'; endif
48.        endfor
49:       if (f" < f*) then            // if a new best solution
50.         S*:= S"                    // update S*
51.       endif
52.       T := T";                     // best ELS solution becomes the new initial solution
53.     endwhile
54.   endfor
55. end
```

*Algorithm 1.* GRASP×ELS *framework for the HVRP*

## 3 Greedy Split procedure

### 3.1 Principle

The Split procedure consists in building an auxiliary acyclic graph $H$ based on a sequence $\lambda$ of tasks (giant tour of customers). The auxiliary graph is composed of $n+1$ nodes numbered from $0$ to $n$ where an arc from node $i$ to $j$, if it exists, represents a subsequence $\mu_{ij}$ of $\lambda$ with $\mu_{ij} = (\lambda(i+1),...,\lambda(j))$ and $Q(\mu_{ij})$ the quantity to collect during trip $\mu_{ij}$. The trip consists in routing from a depot (node number 0) to node $\lambda(i+1)$, from $\lambda(i+1)$ to $\lambda(i+2)$ and so on until $\lambda(j)$ before coming back to the departure depot node. The problem constraints must be satisfied to add the arc (trip) on the graph. Depending on the problem to solve, it is possible to assume that the least-cost paths between tasks have been pre-computed taking into account extra constraints including time-windows on tasks for example. The optimal splitting of $\lambda$ corresponds to a min-cost path from node 0 to node $n$ in $H$. When several resource constraints have to be taken into account, the problem becomes a shortest path problem with resource constraints, as in the HVRP where several vehicles are available. In this case, (Prins, 2009) provided a full description of the Split procedure. The key features are presented below. They include but are not limited to the following keys: label definition, label dominance rule and label generation on node. Note that the arcs of the graph do not need to be explicitly defined at the beginning of the Split procedure but can be generated during the exploration of the shortest path.

Since the HVRP encompasses constraints on resources (number of available vehicles in each type of vehicle), a label is not only a cost to reach a node of the graph but a vector made of a cost plus the availability of resources. A label $L$ is associated to a node and is composed of:

- a solution cost $L.C$;
- for each type of vehicle $k$, the number of vehicles available: $L.a_k$.

It is possible to state that $L = (C, a_1,..., a_K)$ dominates $P = (C, a_1,..., a_K)$ if and only if the following condition holds:

---

$L.C < P.C$ and $\forall k \in [1..K], L.a_k \geq P.a_k$

or

$L.C \leq P.C$ and $\exists k \in [1..K], L.a_k > P.a_k$

---

Thanks to the previous dominance rule, only non dominated labels are saved on nodes. For each subsequence $\mu_{ij} = (\lambda(i+1),...,\lambda(j))$, a cost $h(\mu_{ij}, k)$ is computed assuming trip $\mu_{ij}$ is assigned to one vehicle of type $k$. Vehicles can differ from both fixed cost $f_k$ and cost per distance unit $v_k$ (variable cost).

One can note $L_u^{(i)}$ the $u^{th}$ label saved on node $(i)$ and $\otimes_k$ the operator which applied one trip $\mu_{ij}$ to one label $L_u^{(i)}$ providing a new label $L_v^{(j)}$ on node $(j)$: $L_v^{(j)} = L_v^{(j)} \otimes_k \mu_{ij}$.

The new label $L_v^{(j)}$ is defined as follow: $L_v^{(j)}.C = L_u^{(i)}.C + h(\mu_{ij}, k)$ and $L_v^{(j)}.a_k = L_u^{(i)}.a_k - 1$. $L_v^{(j)} \otimes_k \mu_{ij} = null$ if the generation of a new label using $L_v^{(j)}$ and $\mu_{ij}$ is not possible. This situation occurs when $L_u^{(i)}.a_k = 0$ or when $Q(\mu_{ij}) > Q_k$ i.e. the quantity $Q(\mu_{ij})$ collected exceeds the vehicle capacity.

The dominance rule limits the number of labels saved on each node to the minimal subset and the quantity $Q(\mu_{ij})$ limits the label generation in the graph. However, several authors have previously confirmed that a large number of labels could be saved on node and the total number of labels generated during split process can be huge. A wide spread approach consists in limiting the maximal number of labels generated during Split process and simultaneously the maximal number of labels saved per nodes. They are two parameters of the algorithm which contributes to a strongly efficient split procedure. Lately this key point has been tackled for the Location Routing Problem by Duhamel, Lacomme, Prins and Prodhon (Duhamel *et al.*, 2010).

Notations:

$N_L$ : the maximal number of labels saved on each node;

$N_{\max}$ : the maximal number of labels generated during the Split algorithm.

$G(i)$ : the number of labels currently saved on node $i$

$LB[i]$: the lower bound for node $i$ *i.e.* tackling node $\lambda(i+1),...,\lambda(n)$

$UB$ : the best known solution on node $n$

The following algorithms 2 and 3 use low level procedures to manage the graph $G$:

```
Add_Label_To_Node (P, G, j, UB, LB)
```

This procedure adds, if accepted, label $P$ on node number $j$ from graph $G$. To check if $P$ can be inserted, the inequality $P.C + LB[i] < UB$ must be verified. In addition, the number of labels kept on a node must not be greater than the maximal number allowed per node ($N_L$). Thus, a list of $N_L$ labels is stored in a decreasing order of their cost. If $P.C$ is smaller than the cost of the last label of the list, it is inserted. Insertion can be achieved in $O(N_L/2)$ on average. Note this procedure returns true if the label is added and false otherwise.

```
Apply_Dominance_to_node (G, P, j);
```
This procedure applies the dominance rule between label $P$ and all labels stored in node number $j$. If $P$ is dominated by one label of node $j$ then procedure returns false which means that label $P$ must not be saved. If not, all labels dominated by $P$ are deleted from the list of labels of node $j$.

```
1. procedure Greedy_Split( λ = (0,...,n+1) );
2. begin
3.    Build the graph G.
4.    L = (0, a_1,..., a_K)
5.    Add_Label (G,0,L)
6.    //an upper bound is computed of the f criteria to minimize
7.    UB := Call Compute_Upper_bound(G);
8.    LB := Call Computer_Lower_Bound(G)
9.    // the label UB is added to the last graph node
10.   call Add_Label_To_Node (L, G, n+1, UB);
11.   Total_label := 0;
12.   for i:=0 to n-1 do
13.     for u:=1 to G(i) do // for all labels on node i
14.        L := L_u^(i)    // label number u on node i
15.        Stop := false; j:=i+1;
16.        while (stop=false) and (j<n) do
17.          stop := true
18.          for k:=1 to K do    // for all type of vehicle
19.             if (L_u^(i).a_k > 0) Then  // one vehicle of type k remains available
20.                P = L ⊗_k μ_ij endif // generation of a new label on node G(j) using vehicle of type k
21.             if ( P ≠ null ) then
24.                res := call Apply_Dominance_to_node (G, P, j);
22.                if (res=true) then
23.                   res := call Add_Label_To_Node (P, G, j, UB, LB);
24.                   if (res=true) then
25.                      stop := false; Total_label := Total_label + 1;
26.                      if (Total_Label= N_max ) then break; endif
27.                   endif
28.                endif
29.             endif
30.          endfor
31.          j:=j+1
32.       endwhile
33.     endfor
34.   endfor
35. end;
36. Retrieve the shortest path
```
*Algorithm 2. Generation of labels within the Greedy split for the HVRP*

Algorithm 2 presents an overview of the generation of labels within the greedy split procedure. The *For* loop step 12 to 34 iterates along nodes in $G$. An initial label with null cost is inserted into node number 0 at the first step of the algorithm. All labels of node $i$ are scanned by the *For* loop (steps 13 to 33). $L := L_u^{(i)}$ is the label number $u$ of node $i$ which is used to generated label to node $j$. In step 15, $j$ is assigned to $i+1$. The *For* loop (steps 18-30) investigates propagation of label $L$ using the $K$ types of vehicles. For each $k \in K$, if one remains available at label $L$ (the condition $(L_u^{(i)}.a_k > 0)$ holds) and if the total quantities to collect does not exceed the capacity $Q_k$, a new label is generated using a vehicle of type $k$ to perform the trip $\mu_{ij}$: $P = L \otimes_k \mu_{ij}$. The procedures `Apply_Dominance_to_node` and `Add_Label_To_Node` are used to investigate insertion of label $P$. If one label is inserted into a node $j$, the boolean *stop* is assigned to false and the next value of $j$ will be investigated in the *While* loop. The *While* loop stops if no label has been added into node $j$ (investigating node $j+1$ is useless) or when $j$ exceeds the number of node $n$.

Note a set of non dominated labels is saved on node $n$ and the algorithm return the cost of the less costly label saved on node $n$. The label is used to retrieve the shortest path using the father label of each label. This last step of the algorithm required that the label father is saved on each label by addition of a couple (father_node, father_label) which permits to identify the father label. The basic algorithm to retrieve the shortest path is as folllow:

$L := L_1^{(n)}$ ; `i:=n; j:=1`
**While** (`i ≠ −1`) **loop**
 `the trip is composed of node` $\lambda(L.Father\_Node)...\lambda(i)$
 `i:=L.father_node;`
 `j:=L.father_label;`
**end loop**

## 3.2 Example

Let us consider a HVRP instances with 3 types of vehicles: Type 1 - $a_1 = 3$ - $Q_1 = 10$; Type 2 - $a_2 = 2$ - $Q_2 = 15$; Type 3 - $a_3 = 1$ - $Q_3 = 5$. For the first type, 3 vehicles are available with a capacity of 10, 2 vehicles of type 2 are available (the vehicle capacity is 15) and there is 1 vehicle of capacity 5.

The fixed vehicle cost is $f_1 = 5$, $f_2 = 10$, $f_3 = 5$. Without lose of generality we assume the variable cost are equal for all vehicle types $\forall k, v_k = 1$. The demands to satisfy are as follows: $d_1 = 5$, $d_2 = 3$, $d_3 = 10$, $d_4 = 4$, $d_5 = 5$. The cost function $c_{ij}$ is introduced in table 1. Note this function is non euclidean and $c_{ij} \neq c_{ji}$.

*Table 1*. Cost function between nodes

|   | 0 | 1 | 2 | 3 | 4 | 5 |
|---|---|---|---|---|---|---|
| **0** | 0 | 3 | 7 | 10 | 4 | 10 |
| **1** | 5 | 0 | 5 | 10 | 3 | 1 |
| **2** | 10 | 10 | 0 | 2 | 10 | 5 |
| **3** | 5 | 2 | 5 | 0 | 4 | 2 |
| **4** | 6 | 5 | 7 | 5 | 0 | 6 |
| **5** | 3 | 10 | 10 | 10 | 2 | 0 |

Let us consider the giant trip $\lambda = (5;2;3;1;4)$. The graph $G$ is composed of 6 nodes numbered from 0 to 5. The initial label $L = (0,3,2,1)$ represents a solution where the cost values 0, where there is 3 vehicles of type 1 available, 2 vehicles of type 2 and 1 vehicle of type 3. The arc from node 0 to 1 represents the trip $\mu_{12} = (0, \lambda(1), 0) = (0,5,0)$. The initial label $L = (0,3,2,1)$ is extended from node 0 to node 1 using the three types of vehicles. The first type (fixed cost 5, capacity 10) generates the label $P = (18,2,2,1)$ since the trip cost is $13 = c_{0,5} + c_{5,0} = 10 + 3$ and the number of vehicle of type 1 available is decreased of one unit. The second type (fixed cost 10, capacity 15) generates the label $P = (23,3,1,1)$; the last type (fixed cost 5, capacity 5) generates the label $P = (18,3,2,0)$. Since the dominance rule does not permit to discard one label, the three ones are saved to node 1 (task 5) providing the 3 available ways to service it as illustrated on figure 2.

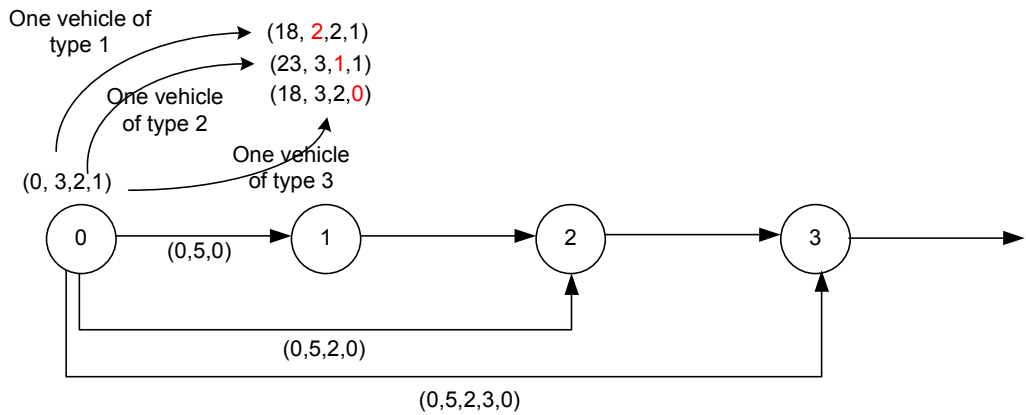

*Figure 2. Label generation from node 0 to node 1*

The arc from node 0 to 2 represents the trip $\mu_{13} = (0, \lambda(1), \lambda(2), 0) = (0,5,2,0)$ with a cost of 30 and total quantity of 8. Thus only vehicle of type 1 and 2 can serve this trip generating 2 labels from node 0: $P = (35,2,2,1)$ and $P' = (40,3,1,1)$.

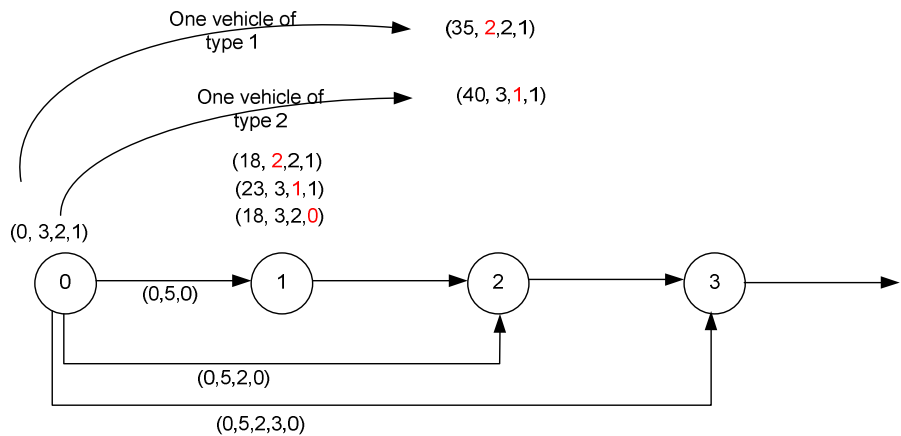

*Figure 3. Label generation from node 0 to node 2*

The arc from node 0 to 3 represents the trip $\mu_{13} = (0, \lambda(1), \lambda(2), \lambda(3), 0) = (0,5,2,3,0)$ with a cost of 27 and total quantity of 18. Since this quantity exceeds the capacity of each vehicle, no label is generated from node 0 to node 3. The process for $i = 0$ stops (the *While* loop of the algorithm stops) and a new step starts by generating labels from node 1.

The arc from node 1 to 2 represents the trip $\mu_{13} = (0, \lambda(2), 0) = (0,2,0)$ with a cost of 17 and total quantity of 3. Each label of node 1 is consecutively scanned and considered to generate 3 labels on node 2 according to the 3 vehicle types. The label $L = (18;2;2;1)$ is extended first using a vehicle of type 1 and gives label $(40;1;2;1)$, second using a vehicle of type 2 (label $(45;2;1;1)$) and lately the using a vehicle of type 3 (label $(40;2;2;0)$).

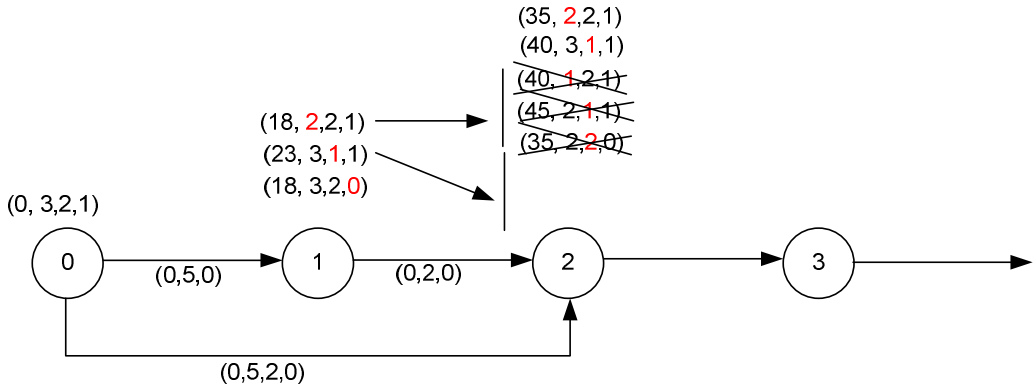

*Figure 4. Label generation from node 1 to node 2*

Note the label $(45;2;1;1)$ is dominated by the label $(40;3;1;1)$ which cost is lower and for which the number of vehicle of each type is greater or equal. In a similar way, let us note label $(35;2;2;0)$ is dominated by label $(35;2;2;1)$. Lastly, $(40;2;1;1)$ is also dominated by $(35;2;1;1)$.

The best label on the last node gives the shortest path in the graph thanks to the father node and the label node. For exemple in figure 5, the father label of $L1$ on node 5 is the label $L3$ on node 3 and the father label of node $L3$ on node $L3$ is the label $L2$ of node 2 etc… To conclude figure 5 gives 3 trips:

Trip 1: $0, \lambda(1), \lambda(2), 0 = depot, 5, 2, depot$

Trip 2: $0, \lambda(3), 0 = depot, ", depot$

Trip 3: $0, \lambda(4), \lambda(5), 0 = depot, 1, 4, depot$

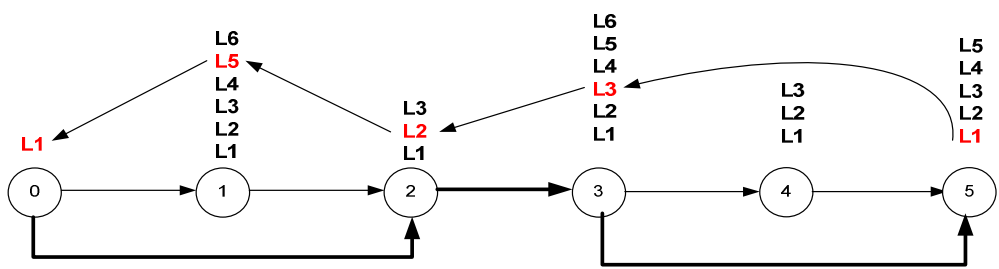

*Figure 5. Example of trips obtained at the end of the split*

# 4 Depth first search Split procedure

## 4.1 Principle

The depth first search Split (DFS Split) procedure (algorithm 3) is designed to first investigate the labels close to the final node trying to reach this ending node as quickly as possible.

While the greedy Split algorithm is managed using a loop scanning all the tasks sequentially from 1 to *n*, the DFS algorithm spreads the labels stored in a stack and iterates until this stack is empty or a maximal number of labels to be generated is reached (lines 11 to 43). The tasks may be visited more than once and backtrackings on the graph imply a fine monitoring of the labels already explored. Thus, in the DFS split, a label is composed of a cost, a n-uplet representing the number of available vehicles of each type and a boolean stating if the label has been previously used during enumeration: $L = (C; a_1, ..., a_K; boolean)$.

The initial stack configuration is composed of the couple $(0; L_0)$ representing the initial label $L_0$ on node 0 with null cost and status *false*. The label to be developed is retrieved by the *pop* procedure (line 12) that removes the couple $(i; L)$ on the top of the stack. Thus, index of the incumbent node is *i* and *current_label* = *L*. A feasible trip $\mu_{ij}$ may generate a label *P* on node *j* depending on vehicle *k* (line 19). *P* is checked to attest that it has not been visited yet (status *false*). If so, the procedures `Apply_Dominance_to_node` and `Add_Label_To_Node` are used to investigate insertion of label *P*. At the end of the *For* loop on the vehicle types (lines 17-34), the new added labels on node *j*, except when *j* = *n* (labels on the last node of the graph cannot be propagated) are considered to form couples $(j; L)$ to save on the top of the stack. This is made by the push procedure (lines 35-39). The next value of *j* (line 40) can then be investigated in the *Repeat* loop until reaching the last node or until $Q(\mu_{ij})$ exceeds the capacities of any vehicle. When a label on node *i* has been spread its status becomes *true* (line 43) and a new one to be developed is selected by the *pop* procedure.

Assume the following algorithms are available for stack management:
```
initialize(P) to initialize stack P
 push(P, L) to add label L on the top of the stack
check_if_empty(P) to test if stack P is empty or not.
 pop(P, L) to remove the last label L on the top of the stack
```

```
1.  procedure Depth_First_Search_Split( λ = (0,..,n+1));
2.  begin
3.    Build the graph G;   L_0 = (0,a_1,...,a_K);   Add_Label (G,0,L_0)
4.    //an upper bound is computed of the f criteria to minimize
5.    UB := Call Compute_Upper_bound(G);
6.    LB := Call Computer_Lower_Bound(G)
7.    Total_label := 0;
8.    call initialize(P);              // initialize stack
9.    call push(P, L)                  // push de initial reference
10.   i:=0;  stop:=false;
11.   while ((check_if_empty(P)=false) and (stop=false) )do
12.    call pop(P, L);
13.    j:=i+1
14.    repeat
15.     Insertion := false;
16.       for k := 1 to K do
17.          if (L.a_k > 0) Then  // one vehicle of type k remains available
18.              P = L ⊗_k μ_{ij}
19.          endif // generation of a new label on node j using vehicle of type k
20.          if ( P ≠ null and P.status = false ) then
21.              Total_label := Total_label+1;
22.              if (Total_Label= N_max ) then break; endif
23.              res := call Apply_Dominance_to_node (G, P, j);
24.              if (res = true) then
25.                 call res = call Add_Label_To_Node (G, j, P, UB, LB);
26.                 if (res=true) then insertion:=true; endif;
27.                 if ((j=n) and (L.cost < BestCost)) then
28.                     Save_G := G;
29.                     BestCost := L.cost;
30.                 endif;
31.              endif
32.          endif
33.       endfor
34.       if (insertion=true) then
35.          for k:=1 to G(j)on node j do
36.              if ( L_k^{(j)}.state = true ) then call push(P, j, L_k^{(j)} ); endif;
37.          endfor
38.       endif
39.       j:=j+1
40.    until ((j>n ) or ( Q(μ_{ij}) > max_{k∈K}(Q_k) ))
41.    current_Label.status := true;
42.  endwhile
43.  Retrieve the shortest path
```
**Algorithm 3.** Generation of labels within the Depth First Search split algorithm

## 4.2 Example

Let us consider the same giant trip $\lambda = (5;2;3;1;4)$ as in the example provided in section 3.2. The graph $G$ is composed of 6 nodes numbered from 0 to 5. The initial label $L = (0,3,2,1)$ creates labels on node 1 and on node 2 using exactly the same principle as the greedy split.

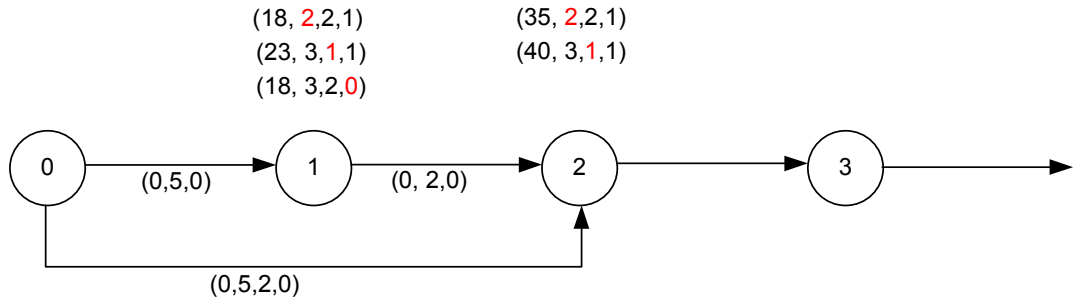

*Figure 6. First step of DFS split*

While the greedy split would then generate labels from the next node of the graph (node 1), the DFS spread labels from the latest generated one, which is $(35;2;2;1)$ on node 2.

The trip $\mu_{23} = (0, \lambda(3), 0) = (0,3,0)$ has a cost of 15 and total quantity of 10. It can be performed using a vehicle of type 1 or 2 only since vehicle of type 3 can not service task 3 which exceeds the vehicle capacity $Q_3 = 5$. Thus label $(35;2;2;1)$ induces two new labels on node 3: label $(55;1;2;1)$ representing assignment of trip $\mu_{23}$ to a vehicle of type 1 and label $(60;2;1;1)$ representing assignment of trip $\mu_{23}$ to a vehicle of type 2.

Then, the trip $\mu_{24} = (0, \lambda(3), \lambda(4), 0) = (0,3,1,0)$ is explored. It has a cost of 17 and total quantity of 14. It can be performed only using a vehicle of type 2. Thus label $(35;2;2;1)$ induces one new label $(62;2;1;1)$ on node 4.

The work on label $(35;2;2;1)$ is now finished since trip $\mu_{25} = (0, \lambda(3), \lambda(4), \lambda(5), 0) = (0,3,1,4,0)$ has a charge of 19, which is greater than the capacity of any vehicle.

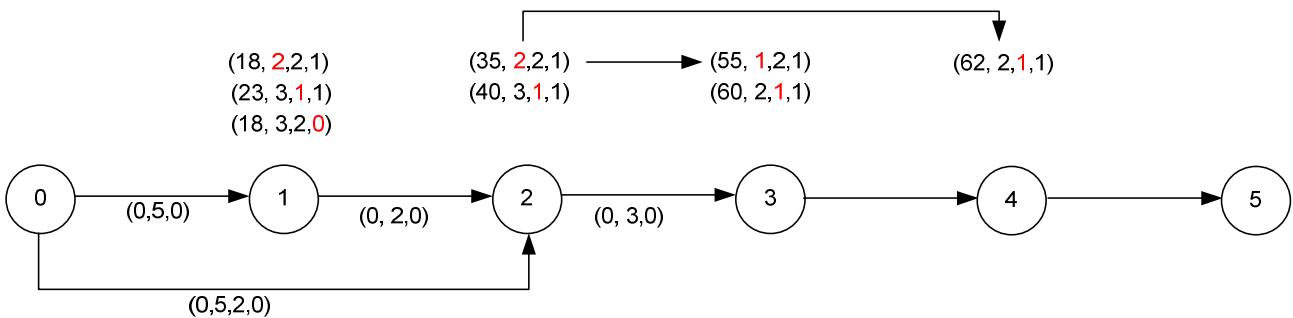

*Figure 7. Second step of DFS split*

The next DFS step consists in extending the label $(62;2;1;1)$ from node 4 to node 5. Backtrack in labels tree is managed using the stack data structure.

# 5 Numerical Evaluation

All procedures in our framework are implemented in Pascal using the Free Pascal compiler. Numerical experiments were carried out on a 2.1 GHz computer running Linux operating system based on an Opteron CPU.

## *5.1 Instances*

**Classical HVRP instances**

Taillard (1999) introduced eight small scale instances with nodes varying from 50 to 100. They are based on VFMP-V files but with limited availabilities of vehicles fleet *i.e.* HVRP. Tarantilis *et al.* (2004) designed a threshold accepting algorithm (TA) for the HVRP and used the eight HVRP instances from Taillard. Li *et al.* (2007) published a record-to-record (RTR) travel metaheuristic for the HVRP that outperforms the TA. Finally, Prins (2009) also tackled these instances with its SMA-D2.

**New Real Life Duhamel-Lacomme-Prodhon_HVRP Instances (DLP_HVRP)**

Using the software introduce by Bajart and Charles (2009) shortest path are available for the 96 French districts using cities with up to 100 citizens or 500 citizens providing instance with 60 to more than 250 nodes.

Shortest paths are computed using the Google web service and represent a true distance in kilometres between cities. The fleet composition has been randomly generated and encompasses instances with up to 10 types. Note that vehicle capacity can be large or very limited: depending on instances, there exist nodes which can not be assigned to some special type of vehicles since they exceed the vehicle capacity type. Instances include subsets where both fixed and variable vehicle costs are not dependant of the vehicles capacity. So it is possible to find instance where the smallest vehicles have the smallest fixed cost and instances where the smallest vehicles are the most costly in terms of both variable and/or fixed cost.

Difficulty arises also in node demand which can exceed the capacity of some type of vehicle, *i.e.* only a subset of vehicles can be well-matched with a node introducing a new difficulty especially for instances where the total fleet capacity is closed to the total demand to service.

To the best of our knowledge, these instances are the first real life instances available based on real country district. They are available at: http://www.isima.fr/~lacomme/hvrp/hvrp.html and at http://prodhonc.free.fr/.
Instances are divided into 4 subsets:
- DLP_HVRP_1: set of 13 small scale instances with less than 100 nodes;
- DLP_HVRP_2: set of 40 medium scale instances with a number of nodes varying from 100 to 150
- DLP_HVRP_3: set of 33 large scale instances with a number of nodes varying from 150 to 200
- DLP_HVRP_4: set of 11 nightmare instances with a number of nodes greater than 200.

## 5.2 Parameters

Trying to favour fair comparative studies, a set of parameters is defined for all classical instances and the new set of instances. The parameters are defined in table2.

*Table 2. Set of parameters for the classical instances and the DLP instances*

|                 | Classical HVRP instances | New Real Life HVRP |
|-----------------|--------------------------|--------------------|
| max_label_total | 10 000                   | 50 000             |
| mlpn            | 3                        | 100                |
| nls             | 500                      | 100                |
| ni              | 10                       | 10                 |
| niv             | 5                        | 5                  |
| nd              | 15                       | 15                 |
| ns              | 15                       | 15                 |
| np              | 50                       | 50                 |
| p               | 1.15                     | 1.15               |

The difference from the two set of parameters is the maximal number of label used during split which is set to 50 000 for the DLP instance since the instances are larger than the classical ones. For a similar reason, the number of label per nodes is increased from 3 to 100.

### 5.2.1 Results on the classical HVRP instances.

Since the best published methods obtain results strongly close to the optimal solutions, it is highly difficult to enlighten significant influence using either the greedy or the depth first search split. However, it is possible to note, the GRASP×ELS with DFS permits to reach the 607.53 value of instance 14 which is not get using the greedy split. This positive result is opposed to the one of instance 20 where the value 1548 does not compete with the SMA-D2 of Prins.

*Table 3. GRASPxELS performances on HVRP instances with DFS Split procedure*

|   |   |   | Taillard |   | Tarantilis |   | Li |   | SMA-D2 |   | GRASPxELS Greedy Split |   | GRASPxELS DFS |   |
|---|---|---|---|---|---|---|---|---|---|---|---|---|---|---|
|   | n | BKS | Cost | time | Cost | time | Cost | time | Cost | time | Cost | time | Cost | time |
| 13 | 50 | 1517.84* | 1518.05 | 473 | 1519.96 | 843 | **1517.84** | 358 | **1517.84** | 33.20 | **1517.84** | 15.36 | **1517.84** | 16.01 |
| 14 | 50 | 607.53* | 615.64 | 575 | 611.39 | 387 | **607.53** | 141 | **607.53** | 37.60 | 609.17 | 24.20 | **607.53** | 71.63 |
| 15 | 50 | 1015.29* | 1016.86 | 335 | **1015.29** | 368 | **1015.29** | 166 | **1015.29** | 6.60 | **1015.29** | 33.86 | **1015.29** | 9.16 |
| 16 | 50 | 1144.94* | 1154.05 | 350 | 1145.52 | 341 | **1144.94** | 188 | **1144.94** | 7.30 | **1144.94** | 22.51 | **1144.94** | 28.19 |
| 17 | 75 | 1061.96* | 1071.79 | 2245 | 1071.01 | 363 | **1061.96** | 216 | 1065.85 | 81.50 | 1065.20 | 23.61 | 1065.20 | 43.12 |
| 18 | 75 | 1823.58* | 1870.16 | 2876 | 1846.35 | 971 | **1823.58** | 366 | **1823.58** | 190.60 | **1823.58** | 129.45 | **1823.58** | 7.24 |
| 19 | 100 | 1117.51 | **1117.51** | 5833 | 1123.83 | 428 | 1120.34 | 404 | 1120.34 | 177.80 | 1120.34 | 144.30 | 1120.34 | 76.44 |
| 20 | 100 | 1534.17* | 1559.77 | 3402 | 1556.35 | 1156 | **1534.17** | 447 | **1534.17** | 223.30 | **1534.17** | 60.33 | 1548.89 | 221.43 |
| Avg. Dev. |   |   | 0.931 |   | 0.617 |   | 0.032 |   | 0.077 |   | 0.10 |   | 0.19 |   |
| Avg Time |   |   |   | 2011.1 |   | 607.1 |   | 285.8 |   | 94.8 |   | 56.70 |   | 59.15 |
| Scale Time - 1.8Ghz |   |   |   | 34.7 |   | 101.7 |   | 213.4 |   | 94.8 |   | 62.37 |   | 65.05 |
| # Best |   |   | 1 |   | 1 |   | 7 |   | 6 |   | 5 |   | 5 |   |

*A GRASPxELS with Depth First Search Split Procedure* for the HVRP ———————————— 20*5.2.2 Results on DLP instances*

Considering simultaneously the 96 instances, it is possible to state that the GRASPxELS based on the DFS split outperforms the GRASPxELS based on the greedy split. Table 4 proves the two GRASPxELS versions have similar average total time and similar average best time (405 and 400). However, the GRASPxELS with DFS split generates 48 best solutions as regards the 36 best ones retrieved by the GRASPxELS with the greedy split.

*Table 4.* DLP_HVRP instances

|  | Average Best Time | # Best | # Equal to |
|---|---|---|---|
| GRASP with Greedy Split | 401.81 | 36 | 12 |
| GRASP with DFS Split | **399.99** | **48** | 12 |

This conclusion must be moderated by a careful analysis of results depending on the 4 set of instances. Indeed, the GRASPxELS based on the DFS split is strongly efficient on instances with less than 150 nodes and the GRASPxELS based on the greedy split is the first ranking methods for larger instances with more than 150 nodes. This is reported on Table 5 where the third line represents the percentage of strictly better solutions retrieved by the methods depending of the subset of instances.

*Table 5.* GRASPxELS performances with the two split procedures for the 4 set of instances

|  | SET_DLP_HVRP_1 | | SET_DLP_HVRP_2 | | SET_DLP_HVRP_3 | | SET_DLP_HVRP_4 | |
|---|---|---|---|---|---|---|---|---|
|  | Greedy Split | DFS Split | Greedy Split | DFS Split | Greedy Split | DFS Split | Greedy Split | DFS Split |
| Best | 2 | **5** | 9 | **26** | **16** | 14 | **9** | 3 |
| equal to | 8 | 8 | 3 | 3 | 1 | 1 | 0 | 0 |
| % of Best | 13.34 | 33.34 | 23.68 | 68.42 | 51.51 | 45.16 | 75.00 | 25.00 |
| Average Best Time | 81.51 | 85.24 | 299.51 | 327.08 | 470.92 | 485.43 | 947.58 | 803.56 |
| Average Gap |  | -0.04 |  | -0.77 |  | 0.79 |  | 0.42 |

In addition, the comparison first based on the number of best solutions found must be round out regarding the average cost and the average computation time as shown on Table 6.

*Table 6.* Average cost of solutions for the 4 set of instances

|  |  | GRASP with Greedy Split | | | GRASP with DFS Split | | |
|---|---|---|---|---|---|---|---|
|  | Number of instances | Average cost | Average computational time | Nb of best | Average cost | Average computational time | Nb of best |
| SET_DLP_HVRP_1 | 15 | 4393.64 | 81.51 | 2 | **4391.34** | 85.24 | **5** |
| SET_DLP_HVRP_2 | 38 | 8704.26 | **299.51** | 9 | **8644.21** | 327.09 | **26** |
| SET_DLP_HVRP_3 | 31 | **11791.08** | **470.92** | **16** | 11875.92 | 485.43 | 14 |
| SET_DLP_HVRP_4 | 12 | **14444.81** | 947.58 | **9** | 14540.93 | **803.56** | 3 |

Except for the last set of instances the computation time for the two GRASPxELS versions are similar. Last line of Table 5 shows per set of instances the relative difference in percent between the costs of each method (greedy version taken as reference). A negative result indicates that the DSF version obtains better solution costs. The conclusion is that the variation between both methods is not strong but clearly encourages the use of one of the versions in relation with the size of the instances.

# 6  Concluding Remarks and Future Research

This article addresses the Split procedure used in routing problems, and provides a new DFS procedure dedicated to the HVRP. It is the second attempt to address in a new way the Split procedure implementation for "non-classic" optimization routing problems where labels encompass both several costs as well as limited resources.

Experiments on the heterogeneous vehicle routing problem prove that DFS Split procedure has a great potential since the GRASPxELS approach using this procedure outperforms the initial GRASPxELS based on the greedy Split. Since the framework used is strictly identical, except for the Split, the improvement depends only on the new procedure. This preliminary study leads us to believe that DFS Split is a promising alternative to the standard greedy Split procedure. Future research is now directed towards new routing problems where the DFS Split could be an efficient way to deal with routing problems. Our research is now oriented towards heterogeneous LRP *i.e.* LRP with heterogonous fleet of vehicles.

# Appendix

**Table A1.** *GRASPxELS performances on French districts- DLP_HVRP_1 ( $n \leq 100$ )*

| Instance name | District name | n | nt | TC (Kg) | TF (Kg) | GRASPxELS with classical split | | GRASPxELS with DFS split | |
|---|---|---|---|---|---|---|---|---|---|
| | | | | | | Cost | time | Cost | time |
| DLP_HVRP_01 | Ain | 92 | 4 | 4000.00 | 7500.00 | **9210.14** | 172.60 | 9219.65 | 71.77 |
| DLP_HVRP_08 | Ardennes | 84 | 3 | 4555.00 | 5000.00 | **4596.52** | 143.30 | 4603.92 | 242.28 |
| DLP_HVRP_10 | Aube | 69 | 4 | 3600.00 | 5350.00 | 2107.55 | 37.80 | 2107.55 | 6.44 |
| DLP_HVRP_11 | Aude | 95 | 4 | 4800.00 | 5750.00 | 3370.52 | 85.84 | **3369.91** | 33.88 |
| DLP_HVRP_36 | Indre | 85 | 6 | 1700.00 | 4650.00 | 5752.79 | 78.81 | **5730.58** | 156.20 |
| DLP_HVRP_39 | Jura | 77 | 5 | 1396.00 | 2500.00 | 2934.55 | 67.54 | 2934.55 | 76.53 |
| DLP_HVRP_43 | Haute Loire | 86 | 7 | 6927.00 | 14450.00 | 8766.54 | 192.60 | **8743.13** | 309.73 |
| DLP_HVRP_52 | Haute Marne | 59 | 3 | 9679.00 | 21500.00 | 4029.42 | 5.09 | 4029.42 | 10.81 |
| DLP_HVRP_55 | Meuse | 56 | 3 | 9484.00 | 11000.00 | 10244.34 | 93.11 | 10244.34 | 117.26 |
| DLP_HVRP_70 | Haute Saone | 77 | 4 | 13755.00 | 16500.00 | 6685.24 | 150.57 | 6685.24 | 87.25 |
| DLP_HVRP_75 | Paris | 20 | 3 | 700.00 | 1150.00 | 452.85 | 0.03 | 452.85 | 0.02 |
| DLP_HVRP_82 | Tarn et Garonne | 79 | 3 | 1900.00 | 3000.00 | 4772.94 | 121.49 | **4768.21** | 114.44 |
| DLP_HVRP_92 | Haut de Seine | 35 | 3 | 5420.00 | 22500.00 | 564.39 | 13.97 | 564.39 | 18.93 |
| DLP_HVRP_93 | Seine saint denis | 40 | 3 | 1900.00 | 3400.00 | 1038.24 | 8.10 | 1038.24 | 10.82 |
| DLP_HVRP_94 | Val de Marne | 47 | 5 | 3475.00 | 8250.00 | 1378.66 | 51.80 | **1378.25** | 22.26 |
| Average | | | | | | 4393.64 | 81.51 | 4391.34 | 85.24 |

*Table A2.* *GRASPxELS performances on French districts DLP_HVRP_2 ( 100 < n ≤ 150 )*

| Instance name | District name | n | nt | TC (Kg) | TF (Kg) | GRASPxELS with classical split | | GRASPxELS with DFS split | |
|---|---|---|---|---|---|---|---|---|---|
| | | | | | | Cost | time | Cost | time |
| DLP_HVRP_03 | Allier | 124 | 4 | 3400.00 | 6000.00 | **10796.00** | 267.13 | 11320.58 | 512.10 |
| DLP_HVRP_05 | Hautes Alpes | 116 | 5 | 6652.00 | 7750.00 | 10969.51 | 276.79 | **10963.62** | 488.63 |
| DLP_HVRP_06 | Alpes Maritimes | 121 | 8 | 6775.00 | 8600.00 | 12153.97 | 168.00 | **11792.94** | 367.91 |
| DLP_HVRP_07 | Ardeche | 108 | 4 | 5365.00 | 7500.00 | **8122.08** | 206.76 | 8130.50 | 306.09 |
| DLP_HVRP_12 | Aveyron | 112 | 4 | 5200.00 | 8100.00 | 3543.99 | 142.31 | 3543.99 | 71.46 |
| DLP_HVRP_13 | Bouches du Rhone | 119 | 5 | 3000.00 | 7500.00 | 6743.92 | 546.75 | **6713.14** | 303.37 |
| DLP_HVRP_16 | Charentes | 129 | 6 | 5700.00 | 8600.00 | 4166.34 | 30.17 | **4161.61** | 180.91 |
| DLP_HVRP_17 | Charentes Maritimes | 105 | 3 | 2000.00 | 3500.00 | 5387.96 | 295.04 | **5370.05** | 172.82 |
| DLP_HVRP_2A | Corse du Sud | 113 | 6 | 2100.00 | 4650.00 | 7932.43 | 214.89 | **7885.93** | 298.92 |
| DLP_HVRP_2B | Haute Corse | 107 | 6 | 1900.00 | 4650.00 | 8719.80 | 227.40 | **8537.31** | 303.14 |
| DLP_HVRP_21 | Cote d'Or | 126 | 3 | 2500.00 | 5500.00 | 5173.56 | 475.95 | **5154.38** | 330.23 |
| DLP_HVRP_25 | Doubs | 143 | 6 | 4100.00 | 8000.00 | 7293.37 | 524.26 | **7228.54** | 518.28 |
| DLP_HVRP_26 | Drome | 126 | 5 | 2700.00 | 10000.00 | **6462.33** | 344.82 | 6481.93 | 350.71 |
| DLP_HVRP_28 | Eure et loire | 141 | 5 | 2400.00 | 6500.00 | 6510.01 | 293.12 | **5542.76** | 343.06 |
| DLP_HVRP_30 | Gard | 112 | 3 | 2100.00 | 7000.00 | 6356.99 | 104.97 | **6321.69** | 201.39 |
| DLP_HVRP_31 | Haute Garonne | 131 | 8 | 2500.00 | 6800.00 | 4132.23 | 312.93 | **4103.88** | 308.39 |
| DLP_HVRP_34 | Herault | 136 | 6 | 2500.00 | 4650.00 | 5891.79 | 373.47 | **5800.12** | 405.62 |
| DLP_HVRP_40 | Landes | 132 | 5 | 2802.00 | 3650.00 | 11256.39 | 471.14 | **11172.98** | 614.92 |
| DLP_HVRP_41 | Loir et Cher | 135 | 7 | 3999.00 | 5850.00 | **7639.10** | 446.15 | 7679.32 | 325.80 |
| DLP_HVRP_47 | Lot et Garonne | 111 | 5 | 18347.00 | 22550.00 | 16238.69 | 91.75 | **16222.94** | 333.85 |
| DLP_HVRP_48 | Lozère | 111 | 5 | 16320.00 | 20550.00 | **21349.59** | 209.31 | 21413.92 | 371.30 |
| DLP_HVRP_51 | Marne | 129 | 3 | 25722.00 | 50500.00 | 7834.20 | 345.97 | **7780.88** | 315.60 |
| DLP_HVRP_53 | Mayenne | 115 | 3 | 17848.00 | 21500.00 | **6459.38** | 152.54 | 6470.49 | 418.17 |
| DLP_HVRP_60 | Oise | 137 | 4 | 23216.00 | 28000.00 | 17079.57 | 243.30 | **17067.85** | 444.32 |
| DLP_HVRP_61 | Orne | 111 | 3 | 17283.00 | 26000.00 | 7302.86 | 153.94 | **7300.10** | 108.21 |
| DLP_HVRP_66 | Pyrénées Orientales | 150 | 4 | 27369.00 | 47500.00 | **12896.15** | 551.11 | 13319.73 | 442.89 |
| DLP_HVRP_68 | Haut Rhin | 125 | 4 | 22254.00 | 30000.00 | **9094.64** | 353.00 | 9135.23 | 269.63 |
| DLP_HVRP_73 | Savoie | 137 | 5 | 4200.00 | 7000.00 | 10882.49 | 595.97 | **10243.66** | 598.34 |
| DLP_HVRP_74 | Haute Savoie | 125 | 5 | 3700.00 | 7000.00 | 12267.63 | 217.71 | **11732.54** | 246.66 |
| DLP_HVRP_79 | Deux Sèvres | 147 | 4 | 3300.00 | 5500.00 | **7282.95** | 368.49 | 7314.89 | 473.69 |
| DLP_HVRP_81 | Tarn | 106 | 4 | 16938.00 | 31000.00 | 10769.99 | 197.44 | **10715.28** | 83.71 |
| DLP_HVRP_83 | Var | 124 | 4 | 2900.00 | 4000.00 | 10036.60 | 315.16 | **10019.83** | 332.47 |
| DLP_HVRP_84 | Vaucluse | 105 | 4 | 2700.00 | 4000.00 | 7271.37 | 234.64 | **7269.55** | 206.41 |
| DLP_HVRP_85 | Vendée | 146 | 4 | 3700.00 | 5000.00 | 8878.18 | 322.98 | **8874.31** | 382.98 |
| DLP_HVRP_87 | Haute Vienne | 108 | 4 | 5100.00 | 8100.00 | 3753.87 | 99.51 | 3753.87 | 104.11 |
| DLP_HVRP_88 | Vosges | 127 | 5 | 7800.00 | 8500.00 | 12562.31 | 958.28 | **12443.41** | 632.22 |
| DLP_HVRP_89 | Yonne | 134 | 5 | 2500.00 | 4500.00 | 7188.97 | 231.59 | **7135.36** | 245.63 |
| DLP_HVRP_90 | Territoire de Belfort | 102 | 4 | 5000.00 | 7100.00 | 2360.83 | 16.77 | 2360.83 | 15.36 |
| Average | | | | | | **8704.26** | **299.51** | **8644.21** | **327.09** |

*Table A3.* GRASPxELS performances on French districts ($150 < n \leq 200$)

| Instance name | District name | n | nt | TC (Kg) | TF (Kg) | GRASPxELS with classical split | | GRASPxELS with DFS split | |
|---|---|---|---|---|---|---|---|---|---|
| | | | | | | Cost | time | Cost | time |
| DLP_HVRP_02 | Aisne | 181 | 4 | 4777.00 | 10000.00 | **11678.44** | 689.81 | 12102.01 | 325.86 |
| DLP_HVRP_04 | Alpes Hautes Provence | 183 | 4 | 3900.00 | 6500.00 | **11030.42** | 667.11 | 11276.45 | 726.38 |
| DLP_HVRP_09 | Ariege | 167 | 5 | 8600.00 | 9750.00 | 7654.45 | 319.39 | **7647.59** | 450.18 |
| DLP_HVRP_14 | Calvados | 176 | 4 | 8000.00 | 10100.00 | **5676.98** | 361.72 | 5679.80 | 448.59 |
| DLP_HVRP_15 | Cantal | 188 | 7 | 8700.00 | 9300.00 | 8367.71 | 905.21 | **8301.63** | 520.82 |
| DLP_HVRP_24 | Dordogne | 163 | 4 | 4700.00 | 6500.00 | 9186.30 | 443.10 | **9183.78** | 609.82 |
| DLP_HVRP_29 | Finistère | 164 | 4 | 2900.00 | 8000.00 | 9176.51 | 122.02 | **9147.39** | 424.95 |
| DLP_HVRP_33 | Gironde | 189 | 7 | 4000.00 | 6650.00 | 9563.18 | 606.39 | **9543.17** | 602.72 |
| DLP_HVRP_35 | Illes et Vilaine | 168 | 6 | 3500.00 | 4650.00 | 9817.94 | 811.07 | **9640.80** | 458.96 |
| DLP_HVRP_37 | Indre et Loire | 161 | 5 | 3000.00 | 6000.00 | 6963.61 | 571.37 | **6921.19** | 383.70 |
| DLP_HVRP_42 | Loire | 178 | 7 | 4666.00 | 8050.00 | **11118.66** | 966.84 | 11713.90 | 316.85 |
| DLP_HVRP_44 | Loire Atlantique | 172 | 3 | 13962.00 | 17000.00 | **12351.49** | 744.39 | 12418.00 | 447.32 |
| DLP_HVRP_45 | Loiret | 170 | 3 | 12561.00 | 16500.00 | 10546.69 | 415.02 | **10519.25** | 450.59 |
| DLP_HVRP_50 | Manche | 187 | 6 | 31519.00 | 60750.00 | 12538.63 | 365.46 | **12508.77** | 646.87 |
| DLP_HVRP_54 | Meurthe et Moselle | 172 | 4 | 28947.00 | 55000.00 | **10426.98** | 565.12 | 11511.62 | 364.47 |
| DLP_HVRP_56 | Morbihan | 153 | 4 | 23325.00 | 26000.00 | **31292.64** | 339.08 | 31292.81 | 394.08 |
| DLP_HVRP_57 | Moselle | 163 | 4 | 26054.00 | 28000.00 | **45112.39** | 471.94 | 45152.42 | 638.93 |
| DLP_HVRP_59 | Nord | 193 | 6 | 36193.00 | 65500.00 | 14367.47 | 476.61 | **14367.14** | 676.23 |
| DLP_HVRP_63 | Puy de Dome | 174 | 5 | 27639.00 | 30500.00 | 20513.10 | 253.10 | **20241.72** | 693.90 |
| DLP_HVRP_64 | Pyrénées Atlantique | 161 | 3 | 26556.00 | 52500.00 | 17157.37 | 70.38 | 17157.37 | 512.03 |
| DLP_HVRP_67 | Bas Rhin | 172 | 5 | 30435.00 | 50000.00 | **11090.66** | 506.65 | 11854.61 | 336.67 |
| DLP_HVRP_69 | Rhone | 152 | 4 | 29800.00 | 35000.00 | **9241.75** | 205.32 | 9276.93 | 508.55 |
| DLP_HVRP_71 | Saone et Loire | 186 | 3 | 6200.00 | 7500.00 | **9936.35** | 389.13 | 9960.84 | 639.69 |
| DLP_HVRP_72 | Sarthe | 186 | 4 | 6100.00 | 9500.00 | **5948.99** | 458.28 | 5976.54 | 197.11 |
| DLP_HVRP_76 | Seine Maritime | 152 | 8 | 5400.00 | 7100.00 | **12086.57** | 426.51 | 12098.66 | 685.64 |
| DLP_HVRP_77 | Seine et Marne | 190 | 3 | 5200.00 | 8000.00 | 7004.97 | 278.69 | **6991.59** | 636.46 |
| DLP_HVRP_78 | Yvelines | 190 | 4 | 5400.00 | 8000.00 | **7066.17** | 439.70 | 7069.82 | 471.38 |
| DLP_HVRP_80 | Sommes | 171 | 3 | 3900.00 | 6000.00 | 6864.75 | 410.38 | **6839.96** | 229.66 |
| DLP_HVRP_86 | Vienne | 153 | 5 | 3900.00 | 6200.00 | 9085.66 | 440.02 | **9076.63** | 383.30 |
| DLP_HVRP_91 | Essonne | 196 | 4 | 9000.00 | 10000.00 | **6419.23** | 672.65 | 6437.14 | 544.07 |
| DLP_HVRP_95 | Val d'Oise | 184 | 2 | 9950.00 | 14000.00 | **6237.61** | 206.09 | 6244.13 | 322.61 |
| Average | | | | | | **11791.09** | 470.92 | 11875.92 | 485.43 |

***Table A4.*** *GRASPxELS performances on French districts DLP_HVRP_4 ( 200 < n )*

| Instance name | District name | n | nt | TC (Kg) | TF (Kg) | GRASPxELS with classical split | | GRASPxELS with DFS split | |
|---|---|---|---|---|---|---|---|---|---|
| | | | | | | Cost | time | Cost | time |
| DLP_HVRP_18 | Cher | 256 | 5 | 5000.00 | 10000.00 | 9797.61 | 1216.10 | **9782.58** | 807.28 |
| DLP_HVRP_19 | Corrèze | 224 | 5 | 4400.00 | 7500.00 | **11805.34** | 1009.87 | 11870.28 | 879.23 |
| DLP_HVRP_22 | Cote d'Armor | 239 | 2 | 6000.00 | 8000.00 | **13162.90** | 835.87 | 13213.34 | 841.87 |
| DLP_HVRP_23 | Creuse | 203 | 4 | 5000.00 | 7000.00 | **7809.20** | 802.30 | 7849.75 | 850.05 |
| DLP_HVRP_27 | Eure | 220 | 5 | 4400.00 | 8500.00 | **8520.74** | 995.85 | 8539.77 | 741.55 |
| DLP_HVRP_32 | Gers | 244 | 8 | 4500.00 | 7650.00 | 9537.48 | 1131.44 | **9402.54** | 1145.63 |
| DLP_HVRP_38 | Isère | 205 | 5 | 3600.00 | 7000.00 | 11439.58 | 421.50 | **11397.37** | 606.97 |
| DLP_HVRP_46 | Lot | 250 | 5 | 40520.00 | 58750.00 | **24805.27** | 1475.05 | 25066.46 | 853.60 |
| DLP_HVRP_49 | Maine et Loire | 246 | 8 | 42700.00 | 59750.00 | **16417.30** | 990.34 | 16569.06 | 706.91 |
| DLP_HVRP_58 | Nièvre | 220 | 6 | 37418.00 | 51000.00 | **23530.10** | 1028.25 | 23826.88 | 751.92 |
| DLP_HVRP_62 | Pas de Calais | 225 | 5 | 42216.00 | 60500.00 | **23434.56** | 828.76 | 23881.43 | 669.64 |
| DLP_HVRP_65 | Hautes Pyrénées | 223 | 3 | 33789.00 | 42500.00 | **13077.63** | 635.64 | 13091.74 | 788.08 |
| Average | | | | | | 14444.81 | 947.58 | 14540.93 | 803.56 |